# Cannabis Impairment Monitoring Using Objective Eye Tracking Analytics

| Author | Affiliation | Contact Information |
| --- | --- | --- |
| Jon Allen | SyncThink | jallen@syncthink.com |
| Leah Brickson | Stanford University | llbricks@stanford.edu |
| Jan van Merkensteijn* | SyncThink | jvmerkensteijn@syncthink.com |
| Dan Beeler | SyncThink | dbeeler@syncthink.com |
| Jamshid Ghajar, MD, PhD, FACS | Stanford University | jghajar@stanford.edu |

Corresponding author is identified with an *

Cannabis Impairment Monitoring Using Objective Eye Tracking Analytics
Jon Allen, Leah Brickson, Jan van Merkensteijn, Dan Beeler, and Jamshid Ghajar

## Table of Contents

# Cannabis Impairment Monitoring Using Objective Eye Tracking Analytics
Jon Allen, Leah Brickson, Jan van Merkensteijn, Dan Beeler, and Jamshid Ghajar

**INTRODUCTION**

With the ever growing legalization of cannabis in the United States, there is an urgent need to be able to rapidly, objectively, and accurately measure impairment resulting from cannabis usage to address public and private safety concerns [1][2]. Traditional pharmo-toxocology measurements provide objective measures but can be highly invasive, require a lengthy time for results processing, and lack a direct measurement of physical impairment.

Currently, the subjective nature of cannabis impairment assessment in the field is a critical deficit in the public safety infrastructure of the United States and throughout private industry. Not only are very few public safety personnel trained to adequately assess impairment using their subjective toolset, but there is also a hierarchy of skill in making these assessments. Regular officers are ill equipped at best, and specialized training in field sobriety or drug recognition are necessary for reliable identification of the impairment type. Additionally, in the occupational safety domain, there are no standardized methods for assessing acute current impairment, only historic impairment as determined by lab testing of urine, blood, or hair. This hole in understanding current impairment level is a threat vector for all industries where attention is at a premium, such as construction or trucking. Even more importantly, this lack of quantified impairment is an even greater threat to the public in occupations where the safety of the general population is dependent on the attention of an individual, such as with commercial airline pilots or clinicians. Finally, from a corporate perspective, quantifying risk using objective metrics allows for a reduction of liability and associated risk premiums paid to insurers.

None of the current approaches for cannabis intoxication detection involve objective data around the extent of impairment. They instead utilize analog and subjective methodologies such as bright light reaction, pupil size evaluation, and manual peripheral nystagmus assessment. We hope that by introducing objective and verifiable metrics into the conversation, the occupational safety, the public safety, and the public communities are all better equipped and better protected against the inherent biases of subjective assessment that may stem from any number of sources including gender, age, race, or ability.

By providing a consistent and predictable .4Hz circular smooth pursuit stimulus in a head mounted display, we can compare performance interindividually and intraindividually to understand expected performance and performance deficits after cannabis consumption. We follow subjects longitudinally by taking baseline measures prior to impairment, and again post-impairment. While other data was collected, such as horizontal saccadic eye movement and simple reaction time tap response, these were not found to be nearly as significant as the variance in smooth pursuit performance and the associated metrics.

# Cannabis Impairment Monitoring Using Objective Eye Tracking Analytics
Jon Allen, Leah Brickson, Jan van Merkensteijn, Dan Beeler, and Jamshid Ghajar

**PHYSICAL MECHANISM OF CANNABIS IMPAIRMENT**

The active components of marijuana are cannabinoids, which bind via cannabinoid receptors at a variety of locations in the body. Within the brain itself, the cerebellum in particular has numerous receptors for the psychoactive cannabinoid THC (CB1). The cerebellum is integral in motor control and fine motor activity. The cerebellum integrates spatial and temporal information to orient attention through prediction, and more specifically directs fine and precise oculomotor and vestibular coordination. While the cerebellum only occupies 10% of the brain's volume, it accounts for over 80% of the brain's total neurons [3][4]. Furthermore, the cerebellum's Purkinje cells have a particularly high concentration of CB1 receptors [5] and also represent 19% of total cerebellar neurons [6], thereby signaling there is a clear centrality of the endocannabinoid system (eCB) in the cerebellar internal signaling structure and methodology. CB1 receptors operate on both presynaptic and postsynaptic sides of the synaptic cleft as well as in astrocytes [7], and when cannabinoids bind to these receptors, the effect is to insulate the neurons, dampening signal transfer (cannabinoids are lipids, so are electrical insulators[8][9]). This mitigates the brain's otherwise more precise response to a stimulus, effectively acting as a low-pass filter [3][4][5][6][7][8][9]. We consider motor control impairment, specifically oculomotor control impairment, to be originating in the cerebellum [10]. By looking at the difference between error distributions for non-impaired and impaired individuals, we can model this impairment reliably with an AUC of .89 for our current model.

**BACKGROUND**

Based on reported observations from Fant, et. al. [11], and subsequent research papers, we determined that impairment measured on a population of 20 individuals would be sufficient to establish significant findings. Additionally, public safety officials have a methodology for differential diagnosis of drug impairments that originated from LAPD's Drug Recognition Expert (DRE) Program. The program relies on manual evaluation of symptoms such as temperature and hyperactivity, and most importantly on ocular symptoms like pupil size, gaze, and nystagmus [12][13].

**STUDY DESIGN**

This study tracks a cohort of healthy individuals before and during cannabis intoxication.

**INCLUSION CRITERIA**

To be included in the study, subjects had to provide informed consent that they would be ingesting cannabis via inhalation, would be conducting eye tracking assessments, would be conducting other cognitive assessments, and that their data would be published and used for commercial purposes. Additionally, all subjects had to be over the age of 21 at the time of cannabis consumption and data collection. Subjects also had to not have any physical attributes that would be prohibitive of use of the EYE-SYNC eye tracking sensor. It is also important to note that subjects with ADHD were included in this study. Subjects with vision correction such as glasses, soft contacts, or LASIK were included in the study but if they utilized glasses or soft contacts, those devices were removed for data collection.

**EXCLUSION CRITERIA**
Subjects were excluded from this study if they were under 21 years of age, were pregnant, or if they wore hard contact lenses. Additionally, subjects with intraocular distance or interpupillary distance (IPD) that was prohibitive of data collection were excluded post data collection.

One volunteer's data could not be gathered consistently due to limitations for ocular distance and size. The following results are for the remaining 19 users.

**DATA QUALITY CONTROLS**
Data collection typically occured between the same hours of 4-8 pm each session to avoid signal noise from circadian rhythm. All subjects had the same, rough, diurnal schedule. Data collection occurred at the same location for most of the subjects. All volunteers were asked not to use cannabis, alcohol, or other impairing substances for at least 24 hours prior to the test. Roughly one third of the volunteers had extensive prior experience using the EYE-SYNC device, to control for learning effects over the course of the tests. A majority of subjects use cannabis frequently (once a week or greater).

**RECRUITING**
Subjects were recruited via the personal networks of the researchers.

**CONSENTING**
Prior to all data collection and dose administration, subjects were walked through the study protocol, expectations, and goals in order to provide them with a comprehensive understanding of the research and their individual tasks. Additionally, all subjects were given clear cautions that they may become impaired by dosing with cannabis. All subjects gave written informed consent for their dosing and subsequent data collection. No pregnant women participated in this study. No minors participated in this study.

**POPULATION**
A group of 20 volunteers was selected from a wide range of demographics and a roughly even distribution of gender and age. No restriction was placed on ADHD assessment, ADHD medication usage, or cannabis usage frequency. No upper bound was placed on the volunteers' age.

Cannabis Impairment Monitoring Using Objective Eye Tracking Analytics
Jon Allen, Leah Brickson, Jan van Merkensteijn, Dan Beeler, and Jamshid Ghajar

Table 01: Population

| Subject ID | Sex | Vision Correction | Vision Correction Type | ADHD | ADHD Medication | Mood | Cannabis Use Cadence | Last Use (Days) |
|---|---|---|---|---|---|---|---|---|
| 01 | M | Yes | LASIK | Yes | 5mg Methylphenidate | Normal | 5x per week | 3 |
| 02 | F | Yes | Farsightedness | No | - | Tired | Weekly | 3 |
| 03 | M | Yes | LASIK | No | - | Good | Daily 3-5x | 0 |
| 04 | F | No | - | No | - | Good | Weekly | 6 |
| 05 | M | Yes | Age related farsightedness | Yes | No | Good/Physically Tired | Weekly | 3 |
| 06 | F | No | - | Yes | No | Great! | Weekly | 8 |
| 07 | F | Yes | Farsighted, stigmatism | No | - | Happy/tired | Daily 3-5x | 3 |
| 08 | M | Yes | Farsighted | No | - | Good/positive | Daily | 2 |
| 09 | M | Yes | Nearsighted | No | - | Content | Daily | 1 |
| 10 | F | Yes | Farsighted, glasses | No | Birth control | Ok | Every other day | 1 |
| 11 | M | No | - | No | - | Chill | Daily | 1 |
| 12 | M | Yes | Glasses | No | - | Contemplative | Yearly | - |
| 13 | F | Yes | Soft contact lenes | No | - | :) | Weekly | 3 |
| 14 | M | No | - | Yes | No | Bitter and jaded | yearly | 121 |
| 15 | M | Yes | Glasses, myopia ~5 diopters | No | - | Skeptical | Daily | 1 |
| 16 | F | Yes | Contact lenses | No | - | Happy | Monthly | 4 |
| 17 | M | No | - | No | - | Jovial | Daily | 2 |
| 18 | M | Yes | Soft contact (-4,-4) | No | - | Elated but a bit tired | Weekly | 1 |
| 19 | M | No | - | No | - | Great! | Daily | 1 |
| 20 | F | No | - | No | - | Good | Daily | 1 |

Cannabis Impairment Monitoring Using Objective Eye Tracking Analytics
Jon Allen, Leah Brickson, Jan van Merkensteijn, Dan Beeler, and Jamshid Ghajar

**PROTOCOL**

Each subject was walked through the procedure prior to the beginning of their particular trial. The procedure itself is broken up into two distinct timepoints:

| | | |
|---|---|---|
| t0 | 0 mins | Baseline measurement |
| t1 | 45 mins | Followup measurement |

Cannabis was administered through inhalation; one dose of .5 grams with 18% THC content was provided. Prior to smoking, each subject completed a baseline that consisted of three iterations of a simple reaction time (SRT) task, three iterations of a horizontal saccade eye movement (saccades) task, and three iterations of a clockwise circular smooth pursuit (smooth pursuit) task. Each round of testing lasted approximately 6 minutes. Three complete tests were required.

SRT, saccades, and smooth pursuit were repeated in triplicate at each of the two timepoints. T1 was chosen at 45 minutes based on the previous research by Fant that suggested a peak in impairment [11].

In addition to the objective data collection using the SyncThink EYE-SYNC eye tracker, each subject was also asked to supply some subjective feedback on their experience. At each time point, t0 and t1, each subject was asked to give feedback on their level of perceived impairment, or "how high" they were. This was recorded by the experiment administrator on the notes section of their demographics form.

**COMPENSATION AND BURDEN REDUCTION**

Subjects were compensated with a small gift bag (approx: $10 value) upon the completion of data collection. Additionally, subjects were offered lunch or dinner upon the completion of their data collection. Finally, to ensure all subjects did not participate in impaired driving, all subjects were driven home via a study-supplied Uber or Lyft.

**EXAMPLE INDIVIDUAL RESULTS**

To highlight the overt impact of cannabis consumption on a single individual, here we provide an exemplar gratis from the study. Below are the raw eye movement traces, relative to the target on the left and relative to the center of the display on the right, as well as the error distributions in both the radial and tangential direction. Herein, we note a substantial change in both error distributions, with the total errors becoming more concentrated and more peaked. In the distributions shown in Fig 3 & 4, the blue curves are the pre-impairment measures and the green

Cannabis Impairment Monitoring Using Objective Eye Tracking Analytics
Jon Allen, Leah Brickson, Jan van Merkensteijn, Dan Beeler, and Jamshid Ghajar

curves are the post-impairment measures. Additionally, in the eye trace plots in Fig 1 & 2, we note an "unraveling" of the traces, indicating a greater degree of variance and jitter.

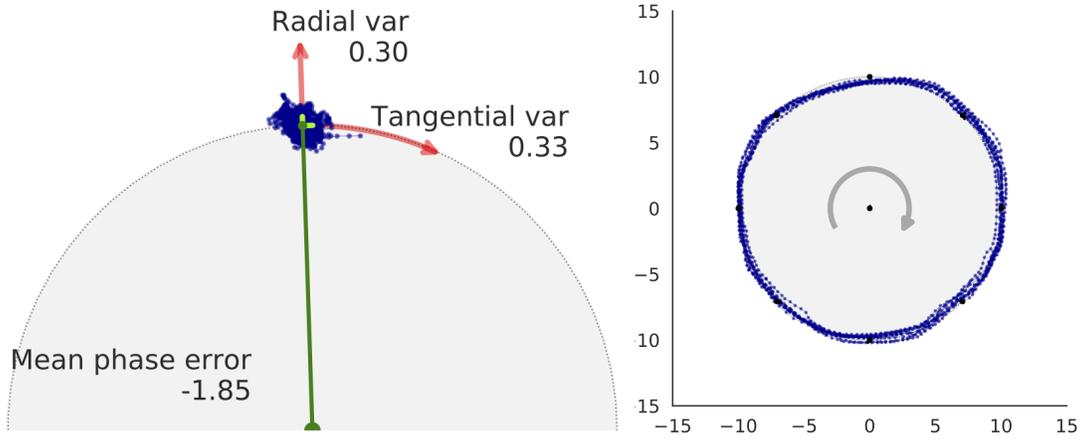

Figure 01: Individual Pre-Impaired Eye Traces

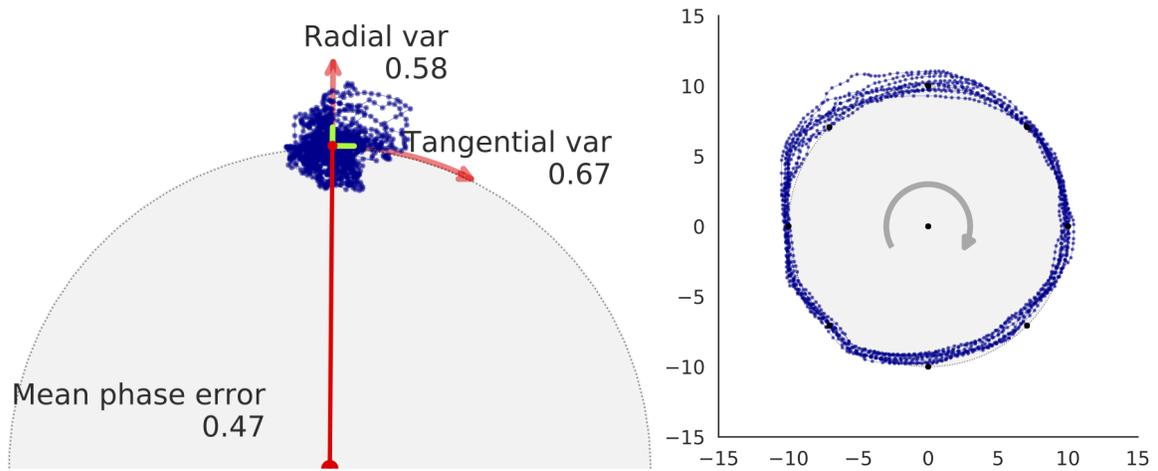

Figure 02: Individual Post-Impaired Eye Traces

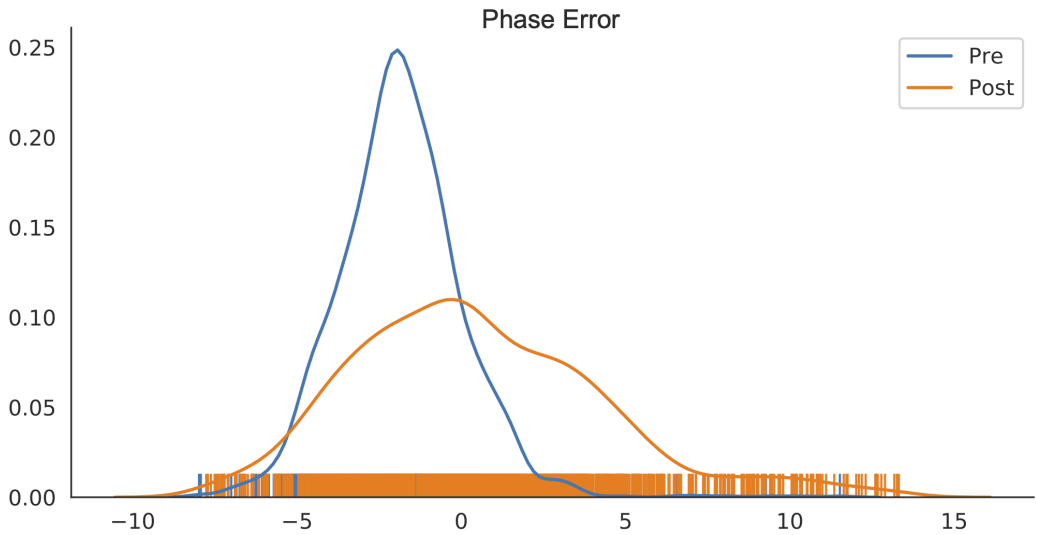

Figure 03: Phase Error Distribution Delta

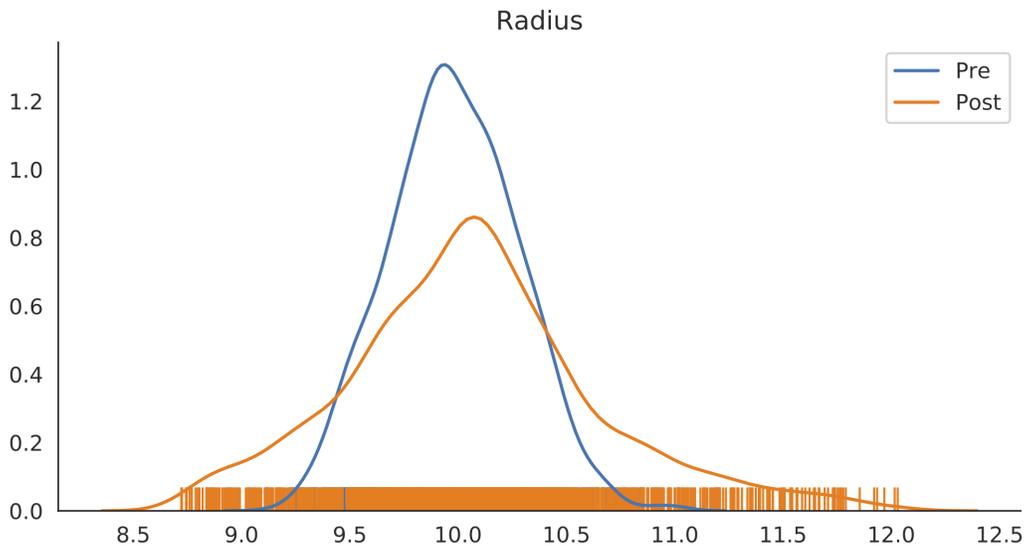

Figure 04: Radial Error Distribution Delta

**GROUP RESULTS**

While we are using a two-tailed, dependent t-test, it would be reasonable to assume that a one-tailed t-test would be appropriate. Thus we are interested in finding all p-values less than 0.05, The following are the relevant metrics. The output is: "Metric name, t-statistic, p-value, Cohen's D effect size, Estimated Number of Observations necessary to establish a p-value of 0.05 at 80% power given the calculated Cohen's D effect size."

Cannabis Impairment Monitoring Using Objective Eye Tracking Analytics
Jon Allen, Leah Brickson, Jan van Merkensteijn, Dan Beeler, and Jamshid Ghajar

Table 02: p-values

| Metric | T-Stat | p value | Cohen's D effect size | Necessary Observations |
|---|---|---|---|---|
| mean radius | -3 | 0.007 | -0.67 | 15.155 |
| V gain | -2.29 | 0.034299 | -0.405 | 39.001 |
| skew radial | -3.10 | 0.006 | -0.728 | 13.1298 |
| skew phase error | 4.82 | 0.0001 | 1.568 | 4.22396 |
| kurtosis phase error | 2.48 | 0.024 | 0.810 | 10.89 |
| blink loss percent | -4.3 | 0.0005 | -1.156 | 6.21961 |

As can be seen from Table 2, there are several markers that produce quite a large effect size and require a small number of observations to produce a statistically significant result. These are the metrics that are particularly good potential candidates for supervised learning models. Taking the mean values per user for both baseline and, separately, impaired metrics, the before and after comparisons for three candidate metrics are presented below in Fig. 5, 6, & 7.

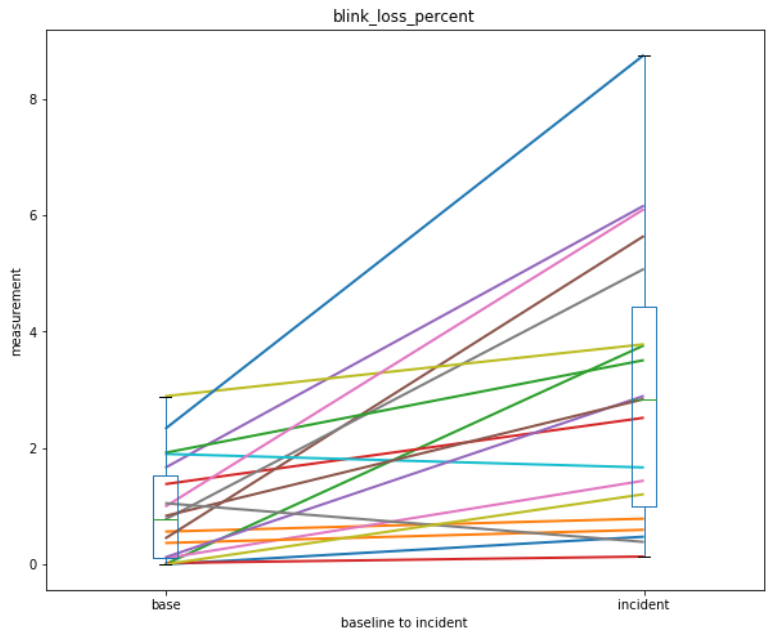

Figure 05: Blink Loss

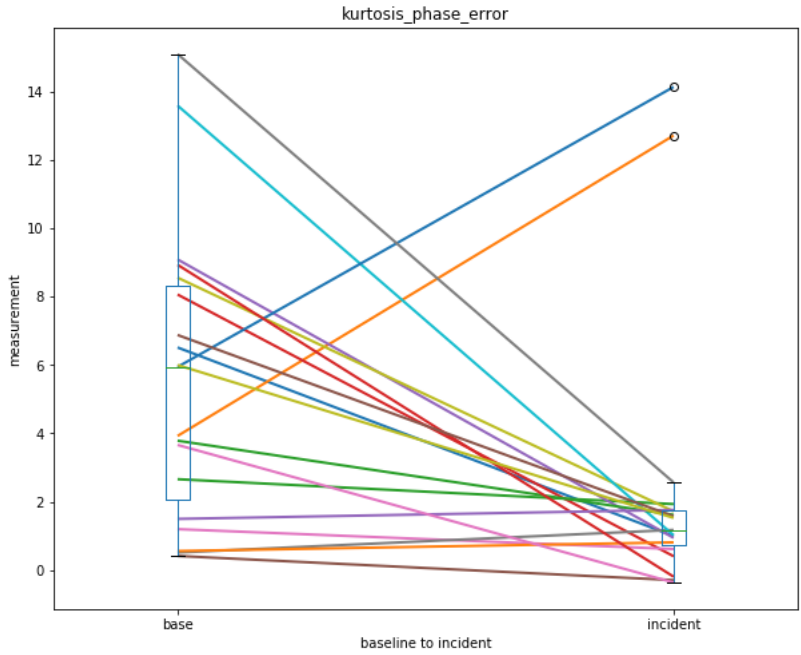

Figure 06: Kurtosis

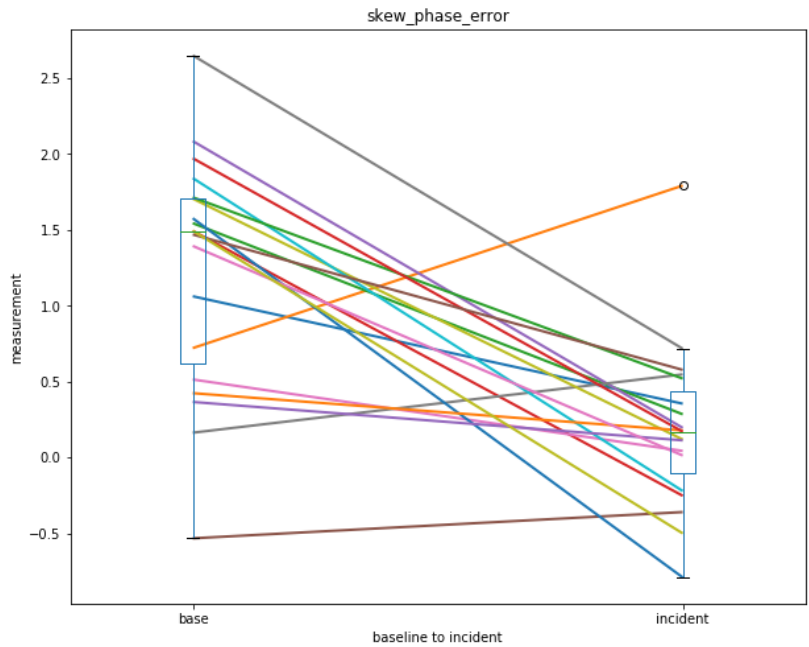

Figure 07: Skew

**SUPERVISED LEARNING**

Using the metrics mean radius, skew radial, skew phase error, kurtosis phase error, & blink loss percentage, a linear classifier support vector machine was constructed using a 50% training vs.

test split for the data set where each individual observation is considered, i.e. for each of the 3 smooth pursuit measurements, the individual metrics are considered rather than the mean. This makes a corpus of 19 ∗ 3 = 57 baseline metrics and 57 impaired metrics for consideration. The test / training split does not group by users.

**WITHOUT PRIOR USER DATA**

We considered a linear model comparing random samples of users not in the trail vs. the impairment observations from the trial. The resulting SVM's had a 75% accuracy rate (ROC median AUC is 0.75, and the best AUC over randomized splits was 0.83). This type of model would be used for law enforcement or anywhere there is a need to test for impairment in a person who has not taken prior data. See Fig. 8

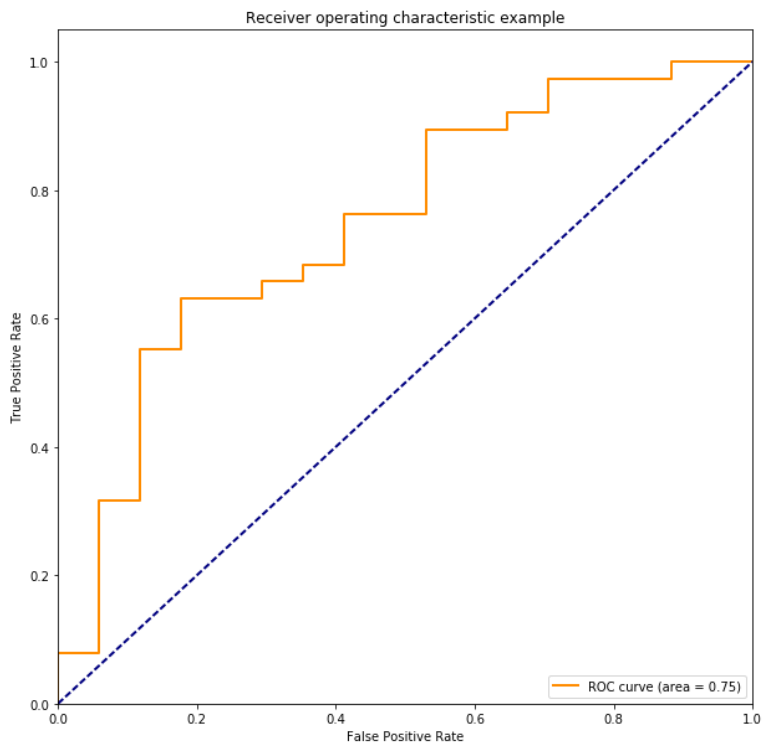

Figure 08: Without Prior User Data

**NORMALIZING USING PRIOR USER DATA**

Lastly, to ascertain the efficacy of inner-user observation, we normalize each individual's raw data by subtracting out the mean of the known sober values from both all initial and all impaired measurements and then comparing the resulting normalized set of volunteers' baseline values with the groups' impaired values. The resulting SVM's has an accuracy of 92% (ROC median

AUC is 0.92, and the best AUC over randomized splits was 0.96). This type of model would be used for users who are tested regularly, e.g. on the work site or at home. See Fig. 9.

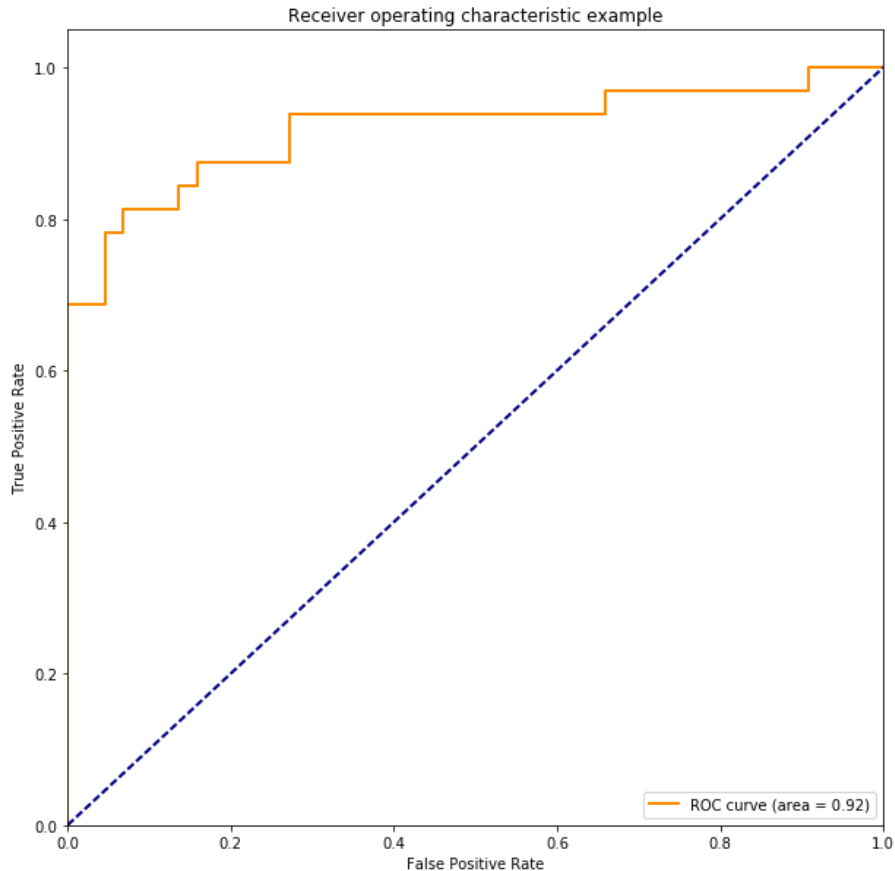

Figure 09: Normalizing Using Prior User Data

**CONCLUSION**

The ability to objectively assess the degree of cannabis impairment intra and inter individually is a hole in public and private safety infrastructure that is best filled by the eye tracking analytics toolkit. By providing portable, rapid, and objective testing, eye tracking can overcome the pitfalls of current objective methodologies such as blood, urine, hair, and breath testing as well as the biases and variable expertise of subjective assessment by individual safety officers. By delivering quantifiable and transparent results in the field, we hope to disrupt the current unscientific status quo with more rigorous and equitable methodologies.

**DATA AVAILABILITY**

The data that support the findings of this study are available from the corresponding author upon reasonable request.

Cannabis Impairment Monitoring Using Objective Eye Tracking Analytics
Jon Allen, Leah Brickson, Jan van Merkensteijn, Dan Beeler, and Jamshid Ghajar

**AUTHOR CONTRIBUTIONS**

| Author | Contributions |
|---|---|
| Jon Allen | - Data collection and subject coordination<br>- Data analysis and model development<br>- Historic literature interpretation and references |
| Leah Brickson | - Data analysis and model verification |
| Jan van Merkensteijn | - Data collection and subject coordination<br>- Manuscript development and editing |
| Dan Beeler | - Study design and coordination |
| Jamshid Ghajar | - Study design oversight<br>- Theoretical analysis of neuromechanics |

**COMPETING INTERESTS**

| Author | Competing Interests |
|---|---|
| Jon Allen | Jon Allen was an employee of SyncThink at the time of the study and was financially compensated for conducting this research. He maintains a financial interest in the company via his equity. |
| Leah Brickson | Leah Brickson has no financial interest in this study. She only stands to gain reputationally through the publication of this research. Both Leah Brickson and Dr. Jamshid Ghajar work at Stanford Medicine. |
| Jan van Merkensteijn | Jan van Merkensteijn is an employee of SyncThink and is compensated for conducting this research. Additionally, he maintains a financial interest in the company via his equity. |
| Dan Beeler | Dan Beeler was an employee of SyncThink at the time of the study and was financially compensated for conducting this research. He maintains a financial interest in the company via his equity. |
| Jamshid Ghajar | Jamshid Ghajar is the founder of SyncThink and is compensated for conducting this research. Additionally, he maintains a financial interest in the company via his equity. |

Cannabis Impairment Monitoring Using Objective Eye Tracking Analytics
Jon Allen, Leah Brickson, Jan van Merkensteijn, Dan Beeler, and Jamshid Ghajar

**REFERENCES**

01  Blows S, Ivers RQ, Connor J, Ameratunga S, Woodward M, Norton R. Marijuana use and car crash injury. Addiction 100, 605-611 (2005).

02  Asbridge M, Hayden J A, Cartwright J L. Acute cannabis consumption and motor vehicle collision risk: systematic review of observational studies and meta-analysis. BMJ 344, e536 (2012).

03  Herculano-Houzel S. The human brain in numbers: a linearly scaled-up primate brain. Front Hum Neurosci 3, 31 (2009).

04  Llinas RR, Walton KD, Lang EJ. Ch. 7 Cerebellum. In Shepherd GM (ed.). *The Synaptic Organization of the Brain*. New York: Oxford University Press (2004).

05  K A Takahashi, D J Linden. Cannabinoid receptor modulation of synapses received by cerebellar Purkinje cells. J Neurophysiol 83(3), 1167-1180 (2000).

06  Nairn JG, Bedi KS, Mayhew TM, Campbell LF. On the number of Purkinje cells in the human cerebellum: unbiased estimates obtained by using the "fractionator". J Comp Neurol 290(4), 527-532 (1989).

07  Busquets-Garcia A, Bains J, Marsicano G. CB1 Receptor Signaling in the Brain: Extracting Specificity from Ubiquity. Neuropsychopharmacology 43(1), 4-20 (2018).

08  Sharma P, Murthy P, Bharath MM. Chemistry, metabolism, and toxicology of cannabis: clinical implications. Iran J Psychiatry 7(4), 149-156 (2012).

09  Nahas G, Harvey D, Sutin K, Turndorf H, & Cancro R. A molecular basis of the therapeutic and psychoactive properties of cannabis ($\Delta^9$-tetrahydrocannabinol). Prog in Neuro-Psychopharmacology & Biological Psychiatry 26, 721-730 (2002).

10  Ghajar J, Irvy RB: The predictive brain state: timing deficiency in traumatic brain injury?. Neurorehabilitation and Neural Repair 22(3), 217-227 (2008).

11  Fant RV, Heishman SJ: Acute and Residual Effects of Marijuana in Humans. Pharmacology Biochemistry and Behavior 60(4), 777-784 (1997).

12  National Highway Traffic Safety Administration (NHTSA) Advanced Roadside Impaired Driving Enforcement (ARIDE) Guidance https://www.nhtsa.gov/enforcement-justice-services/drug-evaluation-and-classification-program-advanced-roadside-impaired

13  National Highway Traffic Safety Administration (NHTSA) Drug Recognition Expert (DRE) Training Guide https://www.nhtsa.gov/enforcement-justice-services/drug-evaluation-and-classification-program-advanced-roadside-impaired

Cannabis Impairment Monitoring Using Objective Eye Tracking Analytics
Jon Allen, Leah Brickson, Jan van Merkensteijn, Dan Beeler, and Jamshid Ghajar

14    Good CH. Endocannabinoid-Dependent Regulation of Feedforward Inhibition in Cerebellar Purkinje Cells. The Journal of Neuroscience 27, 1-3 (2007).

**FIGURES AND TABLES LEGEND**

**Figures**

01    Individual Pre-Impaired Eye Traces

02    Individual Post-Impaired Eye Traces

03    Phase Error Distribution Delta

04    Radial Error Distribution Delta

05    Blink Loss

06    Kurtosis

07    Skew

08    Without Prior User Data

09    Normalizing Using Prior User Data

**Tables**

01    **Population**. A table summarizing the subject population and their demographics.

02    **p-values**. A table summarizing the various p-values of the various metrics.